# Graphs Drawing through Fuzzy Clustering

Mohammadreza Ashouri[1], Ali Golshani[1], Dara Moazzmi[1], Mandana Ghasemi[2]

[1] Faculty of Engineering-Science, University of Tehran
[2] Faculty of Computer Engineering, University of Arak

**Abstract.** Many problems can be presented in an abstract form through a wide range of binary objects and relations which are defined over problem's domain. In these problems, graphical demonstration of defined binary objects and solutions is the most suitable representation approach. In this regard, graph drawing problem discusses the methods for transforming combinatorial graphs to geometrical drawings in order to visualize them. This paper studies the force-directed algorithms and multi-surface techniques for drawing general undirected graphs. Particularly, this research describes force-directed approach to model the drawing of a general graph as a numerical optimization problem. So, it can use rich knowledge which is presented as an established system by the numerical optimization. Moreover, this research proposes the multi-surface approach as an efficient tool for overcoming local minimums in standard force-directed algorithms. Next, we introduce a new method for multi-surface approach based on fuzzy clustering algorithms.

## 1. Introduction

Graphs are used in many scientific fields like computational biology [1] or software engineering [2]. Although graphs theory and algorithms are one of the oldest and most studied area in field of computer sciences, graph drawing problem is rather new. Despite the novelty of studies in this field, the foundation for appearance of graph drawing as a practical art goes back long before development of computer sciences. In all scientific fields, researchers use graphs to show systems which are formed by a large amount of interactive elements, especially when these single elements are simple. For example, electrical engineers draw graphs to represent circuits and social science experts draw graphs for group interactions. However, the most widely uses for graph drawing is in the field of computer and information technology which contains many areas like software architecture [3] or semantic networks [4].

Graph layouts have a considerable impact on time that user requires for understanding its related data. In addition, a graph with poor layout could be confusing and misleading. The purpose of graph drawing is the recognition of the nodes position and edges routing in a manner that clearly shows the structure of the related data.
Generally, there are two types of algorithms for graph drawing. First kind of algorithms focus on special type of graphs like Hamming's graphs and trees. Second type pays attention to general graphs and they are mostly different based on their optimization strategies. This research investigates general graphs and uses a combination of force-directed algorithms and multi-level techniques for drawing them.
Force-directed approach consists of two components, first component is the force or energy model which expresses the quality of drawing and second component is the optimization algorithm for graphical computation which is optimized locally with respect to this model.

On the other hand, multi-level approach is a very useful intuitive tool which we are utilized to overcome the problem of local minimums in standard force-directed algorithms. In general, multi-level algorithms are based on two phases. First in coarsening phase, a group of large graphs are calculated and evaluated in decreased sizes. The second phase is the refinement step, subsequent drawings of more precise graphs are calculated based on drawings of next larger graphs and a suitable kind of force-directed algorithm [5].

Most of the algorithms generate drawings in R² space which consist of isolated nodes and edges which are free-curved lines that connect nodes together. In the most cases, these algorithms assume that input graphs are connected because the computation of connected components and their separate drawing is not difficult [6].

In a wide scope, it is difficult to draw general graphs and the main problem is extra freedom. If a structure is implemented on a graph, practical techniques appear. For example, if we are looking for a graph which is a directed flow we can apply patterns from Sugiyama's shape [7]. On the other hand, we can limit the formation in a proper manner to turn problem into a simpler case [8]. But without these constraints, there is no any simple algorithm for efficient drawing of general graphs.

The most important approaches which were implemented for general graph drawings are trying to facilitate these problems. The topology-shape metrics approach creates orthogonal drawings of general graphs by prioritizing aesthetic [9], and the force-directed approach expresses aesthetic priorities based on forced rules those determine the negative gradient of explicit objective function [10]. The spectral layout technique had been developed based on this observation [11] that if a graph planned in higher dimensional space, the contradiction of its aesthetic metrics is resolved easier and finally multi-level technique has been proposed for overcoming the local minimums problem in standard force-directed algorithms [12].

### 1.1. Force-directed Approach

One of the most useful techniques for studying undirected graphs are reality-based physical models. These techniques are originated in Eades [13] and Kruskal [14] researches. Nodes of a graph are considered as physical objects which are related to various inherent or unessential forces. Some of these forces contain information about the edges, especially as an absorbing force between two endpoints of an edge. The main goal would create a layout for the nodes which minimizes the system's energy, or reaching a stable combination that withstands applied forces on the particles. Some techniques like standard steepest descent or discrete iteration can be used for the desired configuration. Force-directed approach was designed based on these principles and because of their flexibility, simplicity and ability for drawing desired layouts, they have a wide range of applications. Various graph drawing systems had been extended based on this algorithm [15]. The flexibility of force-directed approach lets us consider a wide range of limitations. In practice, these techniques are considerably powerful, because of simplicity in applied algorithms the ultimate drawings maintain the symmetric and hierarchical structure of graph and simultaneously provide a logical distribution of nodes. Various researches had been performed on these kind of algorithms [10, 13] which are resulted in the efficient algorithms that can investigate medium-sized graph.

Force-directed algorithms generally formulate the drawing problem as one of the unlimited numerical optimization. These algorithms rely on a physical model which its main aesthetic consideration is similarity between neighborhood in network and drawing. These algorithms quantify their priorities through forced rules which ensure the fulfillment of the objective or energy function [15].

### 1.2. Multi-level Approach

The existence of many local minimums in the physical model is a major restrictive factor in drawing of large graphs based on standard force-directed algorithms [16]. A system which starts from a random configuration probably involves in a local minimum situation. This situation might a little bit reduces through various iterations of the algorithm. However, applying standard force-directed algorithms for reaching a proper layout in very large graphs is not practically possible. Multi-level technique was proposed for dominating this restrictive factor. This idea was successfully used in many areas such as graph segmentation, and it has been found as a perfect solution for the local minimum problem of the algorithms [17].

## 2. Proposed approach

### 2.1. Force-directed Approach

This study represents force-directed approach for the model drawing of a general graph as a numerical optimization problem. The proposed approach consists of two components: the first component is force or energy model which expresses the quality of the drawing and the second component is an optimization algorithm for the graphical computation which is locally optimized with respect to this model. In proposed model the energy function is not presented explicitly; instead the force rules generate a vector that is the negative gradient of implied energy function which its minimization is the main goal of the approach. Having a clear definition of energy function encourages us to utilize from deep and extended knowledge which is presented by numerical optimization as an established system. Otherwise, using the force rules have some advantages as following [18]:

- Some of the efficient methods are available for energy minimization, without requiring a clear definition of energy function and it extracted only through its negative gradient.
- Using force-based rules give us a great flexibility. Reaching this flexibility based on explicit energy function requires the definition of very intricate functions which intervene in the process of minimization.
- Through having the energy function, its gradient is available too. As a result, besides the efficiency of the algorithm, it is possible to get the optimal layout for an energy function based on the force-directed rules.

The final objective function consists of the linear consensus of energy functions which are corresponding to the force rules. The proposed force rules are raised from two different perspectives: first one is the force rules on the basis of aesthetic metrics and the other one is the rules derived from specific energy functions so that induce interesting visual characteristics.

### 2.1.1 Springs

Springs operate based on Kamada-Kawai model [19]. Kamada-Kawai used a model which relies solely on springs. Their springs follow Hooke's law. Free length and strength of any spring depends on the length of shortest path between nodes of a graph. If the shortest path between u and v nodes have an edge's length of $d(u,v)$ then the spring's free length would be proportional to $d(u,v)$. While, its strength is proportional to $\frac{1}{d(u,v)^2}$. In other words, the amount of the spring force from every pair of different nodes $(u,v)$ applies to each other is proportional to $\frac{1}{d(u,v)^2} |\|X(u) - X(v)\| - K \cdot d(u,v)|$ in which K as a constant shows the desired length of the edge (Fig. 1.). As Kamada and Kawai argued this model would be applied to graphs which have units or weighted edges. In second case the term $(u,v)$ shows the sum of the weights along the shortest connecting path between *u* and *v*.

$$\|f_u\| = \|f_v\| = \frac{1}{d_{uv}^2} \cdot |d - K \cdot d_{uv}|$$

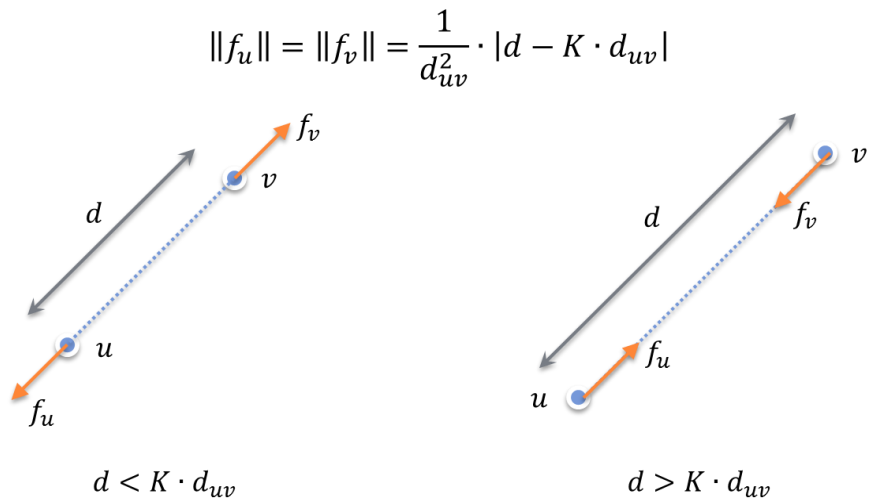

**Fig. 1.** - Springs forces [19]

### 2.1.2 Attraction and repulsion between nodes

According to Fruchterman-Reingold model node-node attraction forces are considered between adjacent nodes in graphs to show the relationships between these nodes [10]. Since the edges determine which pair of nodes are directly connected to each other, the corresponding attraction forces operate as factors that show the importance of these relationships in the drawing. In order to avoid from nodes overlapping and reaching a uniform distribution of nodes in the total area of drawing, they push each other toward outside. The amount of this repulsion force for a specific pair of nodes is a descending function of distance between two nodes.

### 2.1.3 Repulsion between nodes and edges

In the force-directed models, all the forces deal with pairs of nodes. Sometimes these forces allow a node to get too close to an edge. When the edge is short, repulsion between the nodes probably moves the nodes in outward direction of the edges; however, when the edge is long, the repulsion forces cannot avoid great proximity between node and edge because the node can move without nearing to any endpoints of the edge.

Overlapping between node and edge is the most concerned issue. If a node has overlapping with an edge or stay too close to it. It would be difficult to determine if that edge is its implied edge or not. Therefore, we need a short distance and powerful repulsion force between the node and the edge to avoid from this situation. Since we do not use from repulsion forces between the node and the edge for normal distribution of nodes, a long-range force is not necessary. Although, in order to have computational stability, a continuous repulsion force should be used.

The observations show that the repulsion forces between the node and the edge have undesirable effects on forming the structure of graph in early phases. Also, the experiments show that using them only in final phases can fulfill the goals. As a result, the repulsion forces should be considered in the final iterations of the algorithm. Assume that each edge is a contiguous element of very small charged particles that any of its non-neighboring nodes avert edge as a particle with similar electrical charge. Hence, the overall force which is applied to a node is calculated based on infinite sum of forces which are originated from these particles. This idea is the base of defining second degree of repulsion force between the node and the edge.

Consider the node $o$ and the edge $e = (u, v)$ and assume that $q$ is a part of edge $(e)$ with length of $dx$ in the path between $u$ and $v$. This element averts node A with a repulsive force of $df$ which has a direct relation to its length and reverse relation with its distance from A. The direction of this force is along the part and the node. Total applied force over A is defined as an integral over $df$ in the path between u and v (Fig. 2a – 2c.). For simplicity of issue $df$ considered as $\frac{K^\alpha}{x^\alpha} \cdot dx$ and examined it for $\alpha = 2$ and $\alpha = 3$. The conclusion was this fact that $\alpha = 3$ would simplify the outcome of integral because less computational overhead, and the results would be relatively better.

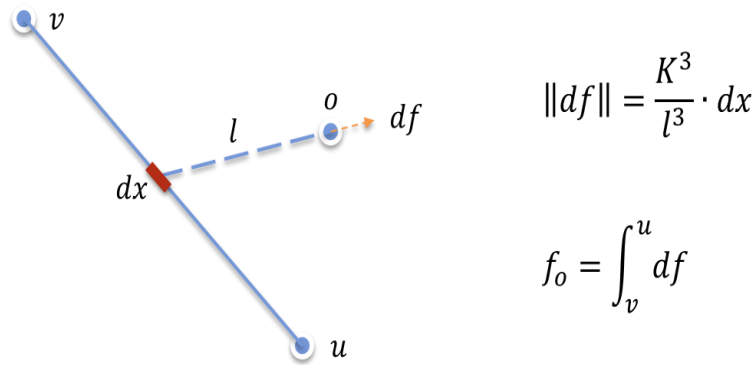

**Fig. 2a.** Repulsive force between the node and the edge [19]

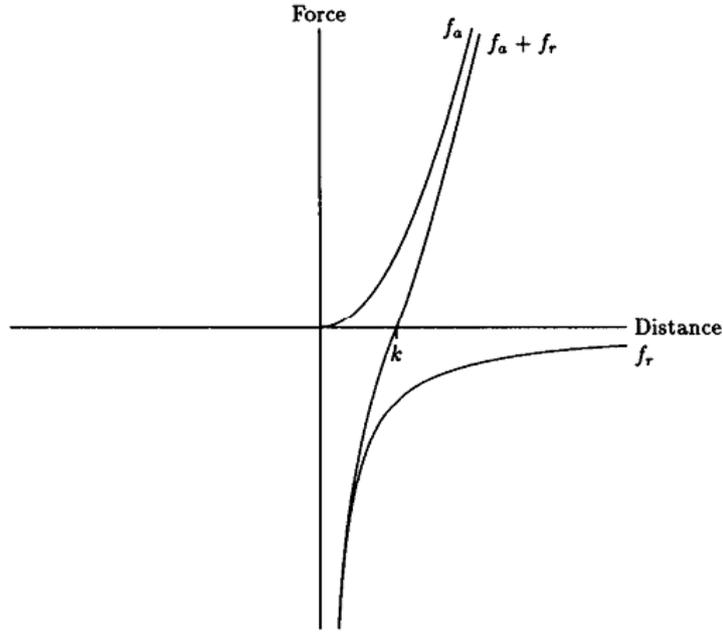

**Fig. 2b.** Forces VS distance in Fruchterman-Reingold Model [19]

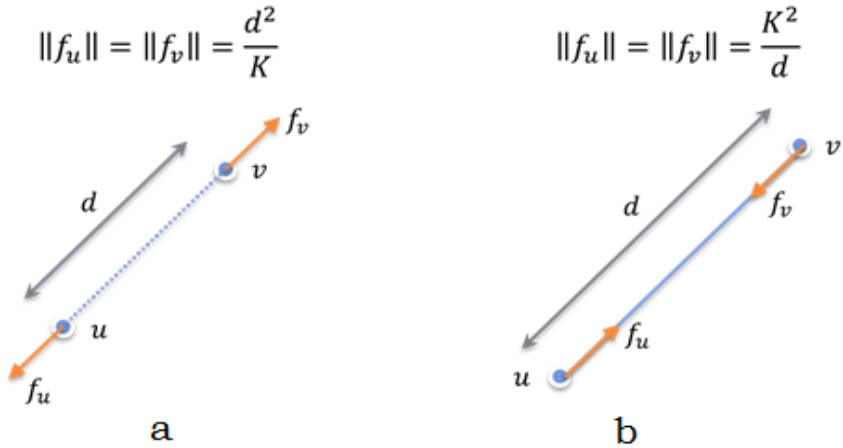

**Fig. 2c.**
a. node-node attraction force [19]     b. node-node repulsive force [19]

### 2.1.4   Stress Function

Stress function would be defined as follow:

$$\sum_{i<j} \omega_{ij}\big(\|X_i - X_j\| - d_{ij}\big)^2$$

Where $d_{ij}$ shows the desired distance between $i_{th}$ and $j_{th}$ nodes, and $X_i$ shows the position of $i_{th}$ node. The normalization constant $\omega_{ij}$ would be equal to $d_{ij}^{-\alpha}$ and often $\alpha = 2$. Desired distance between nodes usually considered as their theoretical distance in the graph (length of shortest distance between nodes). A layout for points which minimizes this function would lead to the best estimation from targeted distances [19].

**Theorem 1:** Assume that $G = (V, E)$ is a connected graph, and $d_{ij}$ is the length of shortest path between $i_{th}$ and $j_{th}$ nodes and $\omega_{ij}$ would be equal to $d_{ij}^{-\alpha}$ for $\alpha > 0$, then $G$ has a layout with stress function of

$$S(X) = \sum_{i<j} \omega_{ij} \left( \|X_i - X_j\| - d_{ij} \right)^2$$

Which would be minimum and it would also minimize this function

$$\sum_{i<j} \omega_{ij} \left( \|X_i - X_j\| - d_{ij} \right)^2$$

$$U(X) = \frac{\sum_{i<j} \omega_{ij} \|X_i - X_j\|^2}{\left( \sum_{i<j} \omega_{ij} d_{ij} \|X_i - X_j\| \right)^2}$$

**Proof:** If the distance between two nodes move toward infinity then the stress function would move toward infinity too. Thus, the distance between nodes in a layout with minimum stress function would be finite; hence, there is a layout with a minimum stress function and finite coordinates. Assume that $p_0$ would be the solution for the following minimization problem:

$$Minimize \quad S(p) = \sum_{i<j} \omega_{ij} \left( \|X_i - X_j\| - d_{ij} \right)^2$$

$p_0$ layout cannot map all nodes to a single position. For understanding this issue assume it is against this situation, which means all nodes in $p_0$ have same position, then

$$S(p_0) = \sum_{i<j} \omega_{ij} d_{ij}^2$$

We can imagine $p_1$ layout like this:

$$X_i = 0, i = 1, \ldots, n-1$$

$$\|X_n - X_1\| = d, 0 < d < 2 \frac{\sum_{i<n} \omega_{in} d_{in}}{\sum_{i<n} \omega_{in}}$$

Then for $p_1$ we would have:

$$S(p_1) = \sum_{i<j<n} \omega_{ij} d_{ij}^2 + \sum_{i<n} \omega_{in} (d - d_{in})^2$$

And consequently

$$S(p_1) - S(p_0) = \sum_{i<n} d \cdot \omega_{in}(d - 2d_{ij}) = d \cdot \left( d \sum_{i<n} \omega_{in} - 2 \sum_{i<n} \omega_{in} d_{in} \right) < 0$$

Which is contrary to the assumption. We define $Q(p)$ function as follow:

$$Q(p) = \sum_{i<j} \omega_{ij} \|X_i - X_j\|^2$$

Through extending the stress function we would have

$$p_0: \quad Minimize \quad \sum_{i<j} \omega_{ij} \|X_i - X_j\|^2 - 2 \sum_{i<j} \omega_{ij} d_{ij} \|X_i - X_j\| + \sum_{i<j} d_{ij}^2$$

The third term is independent of the layout. Thus

$$p_0: \text{Minimize} \sum_{i<j} \omega_{ij}\|X_i - X_j\|^2 - 2\sum_{i<j} \omega_{ij} d_{ij}\|X_i - X_j\|$$

We assume that $c = Q(p_0)$, based on previous result $c > 0$ and $p_0$ layout would be the solution to the below problem

$$p_0: \text{Minimize} \sum_{i<j} \omega_{ij}\|X_i - X_j\|^2 - 2\sum_{i<j} \omega_{ij} d_{ij}\|X_i - X_j\| \text{ subject to } Q(p) = c$$

Which is equal to

$$p_0: \text{Minimize} -2\sum_{i<j} \omega_{ij} d_{ij}\|X_i - X_j\| \text{ subject to } Q(p) = c$$

Since

$$\sum_{i<j} \omega_{ij} d_{ij}\|X_i - X_j\| \geq 0$$

$$\sum_{i<j} \omega_{ij} d_{ij}\|X_i^{p_0} - X_j^{p_0}\| > 0$$

$$p_0: \text{Minimize} \frac{1}{\left(\sum_{i<j} \omega_{ij} d_{ij}\|X_i - X_j\|\right)^2} \text{ subject to } Q(p) = c$$

$$\rightarrow p_0: \text{Minimize } U(p) = \frac{\sum_{i<j} \omega_{ij}\|X_i - X_j\|^2}{\left(\sum_{i<j} \omega_{ij} d_{ij}\|X_i - X_j\|\right)^2} \text{ subject to } Q(p) = c$$

For any $p_1$ layout from $G$ which minimizes $U(p)$ function, we can calculate $p_2 = \sqrt{\frac{c}{Q(p_1)}} \cdot p_1$ with

$$Q(p_2) = \left(\sqrt{\frac{c}{Q(p_1)}}\right)^2 Q(p_1) = c$$

$$U(p_2) = U(p_1)$$

Therefore $p_0$ also would be the solution for the subsequent problem

$$\text{Minimize } U(p) = \frac{\sum_{i<j} \omega_{ij}\|X_i - X_j\|^2}{\left(\sum_{i<j} \omega_{ij} d_{ij}\|X_i - X_j\|\right)^2}$$

**Theorem 2:** Assume a graph $G = (V, E)$, if $p_0$ would be a layout from $G$ with the minimum stress function then:

$$\sum_{i<j} \omega_{ij}\|X_i^{p_0} - X_j^{p_0}\|^2 = \sum_{i<j} \omega_{ij} d_{ij}\|X_i^{p_0} - X_j^{p_0}\|$$

**Proof:** If all coordinates of $p_0$ would be multiplied in the real number $d > 0$, the result stress function of the result layout would be equal to

$$u(d) = \sum_{i<j} \omega_{ij}\bigl(d \cdot \|X_i - X_j\| - d_{ij}\bigr)^2$$

Since $p_0$ is a layout with minimum stress function, the function $u(d)$ would have global minimum at $d = 1$; therefore, $u'(1) = 0$ and

$$u'(d) = 2 \cdot \sum_{i<j} \omega_{ij} \cdot \|X_i - X_j\| \cdot \bigl(d \cdot \|X_i - X_j\| - d_{ij}\bigr) = 0$$

$$0 = u'(1) = 2 \sum_{i<j} \omega_{ij} \cdot \|X_i - X_j\| \cdot \bigl(\|X_i - X_j\| - d_{ij}\bigr)$$

### 2.1.5 The binary stress function

The binary stress function is defined as a linear combination of two functions, and it is used for the computation of the graph layouts [20].

$$B(X) = H(X) + \alpha G(X) = \sum_{(i,j) \in E} \|X_i - X_j\|^2 + \alpha \sum_{i \neq j \in V} \bigl(\|X_i - X_j\| - 1\bigr)^2$$

While the first term relates a layout to structure of a graph through ensuring that the edges are short enough, the second term causes the normal distribution of nodes in a circle (Fig. 3.). The constant $\alpha$ controls the balance between the two terms. The binary stress is suitable for drawing large graphs for two reasons; first for its enhanced scalability and second because of facilitating the desired utilization of space which is crucial for placing a large amount of nodes [20].

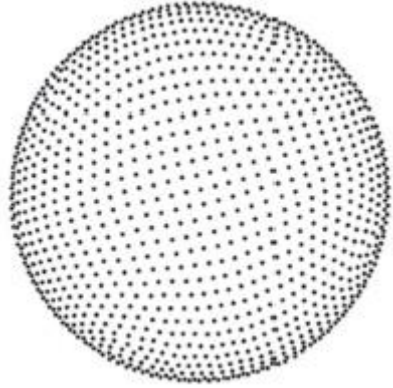

**Fig. 3.** Normal distribution of nodes in the circle [20]

### 2.1.6 Linlog Function

Linlog energy function had been introduced by Andreas Noack as an energy model which induces visual clustering of a graph [21]. The cluster consists of many nodes with a lot of internal edges and a few edges to nodes outside the group. This function would be defined as follow for $p$ layout:

$$U_{LinLog}(p) = \sum_{(i,j) \in E} \|X_i - X_j\| - \sum_{(i,j) \in V^2} \ln\|X_i - X_j\|$$

In a layout from LinLog model with minimum energy, the clusters are clearly separated from the rest of the graph and the distance between every cluster and other parts of the graph is interpretable based on characteristics of the graph.

### 2.1.7 Coefficients

In the definition of the stress function, the produced force from interaction between $i_{th}$ and $j_{th}$ nodes $\left(\|X_i - X_j\| - Kd_{ij}\right)^2$ is accompanied by a coefficient $\omega_{ij}$ which determines the amount of the impact of this force over the stress function. $\omega_{ij}$ usually is defined equal to $d_{ij}^2$ to reflect this issue that the nodes with lower distances in a graph have a greater effect on the final objective function. In the binary stress and Linlog, all forces have a constant coefficient equal to unity. Although the observations show that defining coefficients similar to the stress function can enhance the drawing results in some cases. In the designed program, we make it possible to define the coefficients of this function as a mathematical power of the nodes' distance in the graph.

The following equations show the applied force on the $i_{th}$ node on the force-directed rules

- Spring

$$F_{Spring}(v_i) = -\sum_{j \neq i} \frac{1}{d_{ij}^2} \left(\|X_i - X_j\| - K \cdot d_{ij}\right) \frac{\overrightarrow{X_i - X_j}}{\|X_i - X_j\|}$$

- Head to head attraction

$$F_{Attractive} = -\sum_{(i,j) \in E} \frac{\|X_i - X_j\|}{K} \left(\overrightarrow{X_i - X_j}\right)$$

- Head to head repulsion

$$F_{Repulsive} = \sum_{j \neq i} \frac{K^2}{\|X_i - X_j\|^2} \left(\overrightarrow{X_i - X_j}\right)$$

- Stress function

$$F_{Stress}(v_i) = -\frac{1}{\sum_{j \neq i} \omega_{ij}} \sum_{j \neq i} \omega_{ij} \left(\|X_i - X_j\| - d_{ij} \cdot K\right) \cdot \frac{\overrightarrow{X_i - X_j}}{\|X_i - X_j\|}$$

- LinLog function

$$F_{LinLog}(V_i) = -\sum_{(i,j) \in E} \frac{\overrightarrow{X_i - X_j}}{\|X_i - X_j\|} + \sum_{j \neq i} \frac{\overrightarrow{X_i - X_j}}{\|X_i - X_j\|^2}$$

- Binary stress function

$$F_{BinaryStress}(V_i) = -\sum_{(i,j) \in E} \left(\overrightarrow{X_i - X_j}\right) - \alpha \sum_{j \neq i} \left(\|X_i - X_j\| - K\right) \cdot \frac{\overrightarrow{X_i - X_j}}{\|X_i - X_j\|}$$

$K$ is the desired length of an edge and $d_{ij}$ is the length of shortest path between $i_{th}$ and $j_{th}$ nodes.

### 2.1.8 Optimization Method

The graph drawing problem is considered as a force or an energy model. In previous part the first method is selected and this method described that how aesthetic criteria with the force rules could be summarized. In this regard, it seems appropriate to consider the summation of these force vectors as the negative gradient of an energy function which should be minimized. This implicit energy function is referred as our objective function. The continuous first degree optimization processes are chosen and these processes are repetitive. In every iteration they improve drawing (which is a vector in $R^{2n}$ for two-dimensional drawings) with displacement through a vector $p \in R^{2n}$. The problem of computing this vector is divided into two parts which are the problem of selecting the search direction (direction of $p$) and the problem of step length determination (size of $p$).

- **Computation of forces**

The algorithm requires the computation of the consequent forces which are applied to every node in every iteration of the algorithm. Computing the sum of these forces is a time consuming process that its complexity is $O(|V|^2)$ for head to head driving forces and is $O(|V||E|)$ for repulsive forces between the node and the edge. Their high time-complexity make it difficult to apply the algorithm for large graphs. Different methods had been proposed for reducing the time-complexity of these computations. The Computation of repulsive force between nodes is like n-body problem in physics which is fully studied [22]. Investigating the proposed methods in this domain provide a useful scientific guidance for overcoming computational complexity in this problem. These methods are often extensible to repulsion forces between node and edge.

- **Selecting search direction**

Eades [13] and Fruchterman-Reingold's algorithms used the steepest descent method to determine search direction. They considered the search direction in the same direction of applying the force to each node. Kamada-Kawai's algorithm displaces the node which endures the highest pure force to the local minimum energy point through two dimensional Newton-Raphson technique in every iteration. Our algorithm tries to use from advantages of other methods.
- The force corresponding to each of the force rules will be applied to each of the graph nodes. Hypothetically this force would be equal to the negative gradient of the related energy function. However, for some of simple force rules it is possible that this force being adjusted based on two dimensional Newton-Raphson method.
- For determining search direction, we use the steepest descent in the early iterations and conjugate gradient for the final steps.
- In the steepest descent technique, the position of a node gets updated exactly after computing the applied force to the node and before computing the forces for all nodes, this process enhances the convergence of the algorithm, it is very much similar to this fact that in repetitive linear systems solvers, Gauss-Seidel algorithm is often faster than Jacobi algorithm.

The step length in each iteration is determined based on the search direction algorithm.

- **Step-Length Determining**

Eades' optimization process applies steepest descent technique in its original form. A fixed step-length is the main characteristic of Eades' procedure, and there is no guarantee that this step length would be acceptable, especially when the step-length is too big, the optimization process might fluctuate and never converge to a local minimum. Fruchterman-Reingold's process starts with the calculation of negative gradient and after that, instead of calculation of step-length, it would investigate search's direction components for each node independently in order to confine the maximum distance that a node can replicate its movement on it. This maximum distance is estimated based on temperature which is a descending function of applied iteration until that moment [13].
Cooling scheduling that is used in most of force-directed algorithms makes it possible to generate large displacement at the start of iteration (large step-length), but step-length decreases with progression of the algorithm. Walshaw [12] used a simple scheme in which $temperature = t \cdot temperature$, for $t = 0.9$ would be an ideal solution for multi-level force-directed algorithm. However, it is more efficient to apply an adaptive step-length updating in force-directed algorithms with random primitive conditions in order to prevent from local minimums. This adaptive scheme is originated from trust region algorithm in optimization processes [23]. In this algorithm, step-length can increase or decrease with regard to the progression. The idea of algorithm is that the step-length remains constant if energy decreases. If energy decreases more than five times consequently, then the step-length increases. The step-length decreases only if energy increases (Fig. 4.). Another method is the determination of step-length for each node locally and independently from other nodes. In this technique for each node ($v$) a local temperature variable $heat[v]$ controls the size of displacement vector for that node. The algorithm for calculation of local temperature is shown in Fig. 5. There are three conditions for the calculation of local temperature:

- If $displace[v]$ or $oldDisplace[v]$ be equal to zero vector, the size of $heat[v]$ would not change.
- If $v$ fluctuates around a static position or moves in fixed direction, the local temperature would be updated $heat[v] \cdot (1 + cos \cdot r \cdot s)$ (reduction in first case and increase in second case)
- In all other situations local temperature would be adjusted equal to $heat[v] \cdot (1 + cos \cdot r)$.

```
Function UpdateTemperature (temperature, Energy, Energy⁰)
    if (Energy < Energy⁰)
        progress = progress + 1;
        if (progress ≥ 5)
            progress = 0;
            temperature = temperature / t;
    else
        progress = 0;
        temperature = t · temprature;
```

**Fig. 4.** The step-length decreases by energy increases

Time complexity for updating the local temperature for any node ($v$) is a fixed value; hence, the overall time complexity for the estimation of local temperatures would be linear. It seems that like local temperatures technique is the best choice while applying steepest descent technique and the weakest choice in case of applying conjugate gradient technique. Linear search is another option for specifying step-length in each iteration which has limited usage because of its high cost.

```
Function UpdateLocalTemperature (v)
    if (‖displace[v]‖ ≠ 0 and ‖oldDisplace[v]‖ ≠ 0)
        cos[v] = (displace[v] · oldDisplace[v]) / (‖displace[v]‖ · ‖oldDisplace[v]‖);
        r = 0.15,  s = 3;
        if (oldCos[v] · cos[v] > 0)
            heat[v] = heat[v] × (1 + cos[v] · r · s);
        else
            heat[v] = heat[v] × (1 + cos[v] · r);
        oldCos[v] = Cos[v];
```

**Fig. 5.** Updating the local step-length

### 2.1.9. Summarize Approach

The graph drawing problem is investigated with three independent sub-problems which are mentioned as follows:
- Defining force rules which measure aesthetic metrics.
- Calculation of forces in order to acquire the negative gradient of an implicit energy function
- Using numerical optimization for extracting local minimums for this energy function. In Fig. 6. An overview of final pseudocode is presented.

$$\begin{aligned}
&\textbf{Algorithm } ForceDirectedLayout\ (G)\\
&\quad Initialize\ d_{ij}, K, heat;\\
&\quad \textbf{for } (i = 1\ \textbf{to}\ \alpha \cdot maxIteration)\\
&\quad\quad \textbf{foreeach } (v \in V)\\
&\quad\quad\quad CalculateForce\ (v);\\
&\quad\quad\quad displace[v] = SteepestDescents[v];\\
&\quad\quad\quad UpdateLocalTemperature\ (v);\\
&\quad\quad\quad displace[v] = heat[v] \cdot \frac{f_v}{\|f_v\|};\\
&\quad\quad\quad X_v = X_v + displace[v];\\
&\quad\quad Copy\ (displace, oldDisplace);\\
&\quad \textbf{for } (i = 1\ \textbf{to}\ (1 - \alpha) \cdot maxIteration)\\
&\quad\quad CalculateForces();\\
&\quad\quad ConjugateGradient();\\
&\quad\quad UpdateTemperature\ ();\\
&\quad\quad \textbf{foreeach } (v \in V)\\
&\quad\quad\quad displace[v] = ConjugateGradient[v];\\
&\quad\quad\quad displace[v] = temperature \cdot \frac{f_v}{\arg\max_u \|f_u\|};\\
&\quad\quad\quad X_v = X_v + displace[v];
\end{aligned}$$

**Fig. 6.** Force-directed algorithm

### 2.2 Multi-Level Approach

Although a long range force estimation through appropriate data structure decreases the complexity of force-directed algorithms significantly, it adjusts one node instead of shaping the whole area in every iteration. Large graphs usually have many local minimum energy configurations and this algorithm might probably be used for resolving one of these local minimums. Multi-level approach had been used in many combinatorial optimization problems such as graph clustering [17,24,25] matrix ordering [26] and travelling salesman problem [27]. It proved that it is a useful meta-intuitive tool [28]. Multi-level approach had been also used for graph drawing [24,29,30]. Multi-level approach has three different phases: graph coarsening, initial layout, and layout interpolation [12]. In graph coarsening phase a group of large and larger graphs are formed like $G^0, G^1, \ldots, G^l$. In every larger graph $G^{k+1}$ the goal is the availability of required information for the layout of its parent graph $G^k$ which has less nodes and edges. Then graph coarsening would be continued until a graph with a few nodes acquired. Optimal layout for the largest graph can be identified easily. The layout of larger graphs moves toward narrower graphs recursively with more refinements in every level of graph (Fig. 7. Fig. 8.).

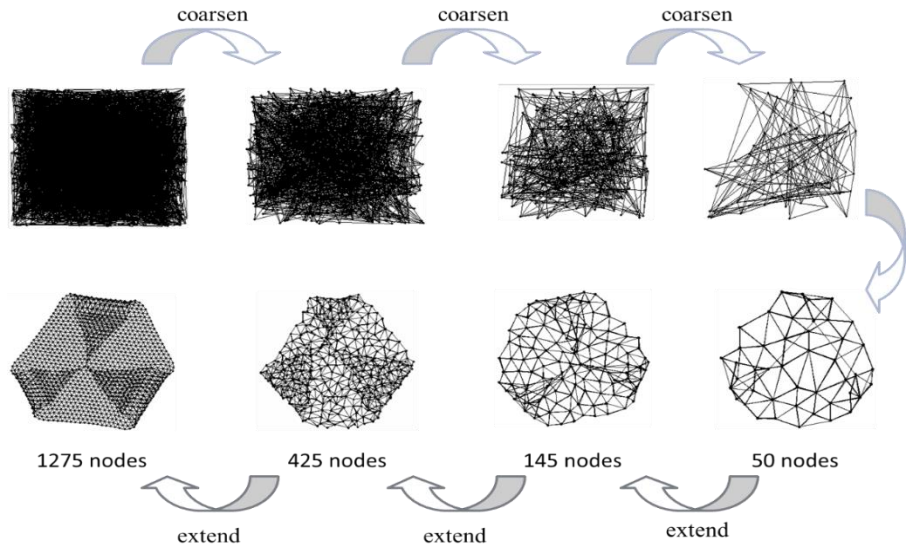

**Fig. 7.** Multi-Level Technique

```
Algorithm MultilevelLayout (G)
    Initialization;
    Graph G^0 = G;
    i = 0;
    while (|V^i| ≥ threshold)
        Graph G^{i+1} = CoarsenGraph(G^i);
        i = i + 1;
    while (i ≥ 0)
        ComputeLayout(G^i);
        if (i ≥ 1)
            InterpolateInitialPositions(G^{i-1});
        i = i - 1;
    return G^0;
```

**Fig. 8.** Multi-Level Algorithm

### 2.2.1 Graph Coarsening

There are several ways for undirected graphs coarsening. One of the frequent ways is based on edge collapsing (EC) [17,24,25] in which adjacent nodes pairs are selected and each pair get combined in a new node. Each node of produced larger graph has a dependent weight which is equal to its main nodes. Collapsing edges are usually selected through maximal matching. Maximal matching is a maximum set of edges that none of its pairs have the same coinciding node. The priority of edges selection - while searching in neighbors list for finding unmatched nodes - has an impact on the quality of results. The simplest choice is the selection of first available edge and another option can be the random selection of objective edge. However, prioritizing in a proper manner can improve the quality of generated layouts significantly. Here, these priorities can be considered:
- Matching with thick edges: here the goal is preferably the collapse of thicker edges, in the process of searching for unmatched nodes in neighbors list, the edge with highest weight would be selected.

- Low weight nodes matching: preserving the balance in nodes' weights through matching between adjacent nodes with lowest node weight.

- Matching based on number of common neighborhoods: for nodes $u$ & $v$ the semi-distance $d^2(u,v)$ is defined based on:

$$d^2(u,v) = 1 - 2 \cdot \frac{|N_u \cap N_v|}{|N_u \cup N_v|}, \qquad N_u = \{u|(u,v) \in E\} \cup \{u\}$$

The priority for matching can be nodes with less semi-distance

Furthermore, other coarsening techniques had been proposed. In [31] maximal independent vertex set (MIVS) had been selected as the nodes of the larger graph. Independent vertex set is a sub-set of the nodes (vertexes) that none of its pairs are beside each other. This independent set would be maximum in conditions where adding another node always leads to dependency. If the distance between two nodes would not be more than 3, then the edges of the larger graph could be generated by connecting them in MIVS through an edge. Fig. 9. Shows a graph (left) and the result of the graph coarsening with EC (middle) and MIVS (right).

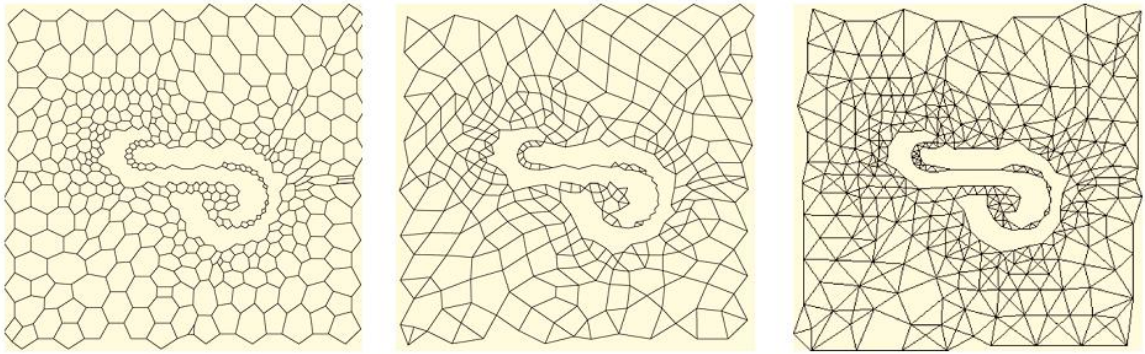

**Fig. 9.** Coarsening process for a graph with 788 nodes(left), large graph produced by EC algorithm with 403 nodes (middle), large graph produced with MIVS with 322 nodes (right) [31]

Graph coarsening phase could be done through clustering and partitioning algorithms. In these algorithms the nodes of graph are partitioned to separate sub-sets that their combination results in generation of larger graph. The weights for nodes and edges in produced graph would be define similar to what has been described in EC technique. While in clustering the number of clusters is determined based on graph structure and through an algorithm, in partitioning the number of partitions is determined based on input parameters. In order to have an efficient multi-level approach, it is necessary that the number of nodes in the result graph would be half of the nodes in the source graph.

*2.2.2 Refinement Phase*

The layout of larger graphs moves recursively toward narrower graphs with more refinements in each level. If graph $G^{i+1} = (V^{i+1}, E^{i+1})$ is derived from $G^i = (V^i, E^i)$ through edge collapsing (EC), two $u$, $v$ nodes which belong to $u \in V^{i+1}$ would be reduced to $u \in V^{i+1}$ and hold the position of $u$ node. If $G^{i+1}$ is derived from $G^i$ through MIVS, then $u \in V^i$ would inherit its position from $G^{i+1}$ or $v$ should have one or more than one vicinity in MIVS, and in this case the position of $v$ would be adjusted to average positions of adjacent nodes.
After this $G^{i+1}$ would take its primary layout which would be refined with force-directed algorithm. If primary layout of two nodes are in a similar position, random displacement would be applied to separate them. Since, primary layout is derived from the layout of the larger graph and this layout generally has a superior position in all coordinates and only some local adjustments would be needed.

Although the primary layout of $G^i$ graph which is extended from $G^{i+1}$ graph usually has an appropriate position in all situations, but the simple implementation of force-directed algorithm can increase the displacements of nodes. Thus, useful acquired information would potentially be removed. For example, in case of spring model, physical distance between two nodes $u$, $v$ in primary layout in $G^i$ is almost similar to distance between corresponding nodes in $G^{i+1}$ graph which means:

$$\|X_u^i - X_v^i\| \approx d_{G^{i+1}}(u', v')$$

Here $u'$ and $v'$ are two nodes in larger graph $G^{i+1}$ which are corresponded to $u$ and $v$. Nonetheless for energy minimization at level $i$ should be as follows:

$$\|X_u^i - X_v^i\| \approx d_{G^i}(u, v).$$

Usually $d_{G^{i+1}}(u', v')$ is much smaller than $d_{G^i}(u, v)$. If the primary layout be used without any changes, then the graph should be extended through a set of large and important displacement for calculation of minimum energy which is inefficient and information about desired primary layout might be lost. As a solution to this problem, early coordinates could be scaled based on the proportion of diagonals between two posterior graphs in Multi-Level algorithm based on spring model.

$$\gamma = \frac{diam(G^i)}{diam(G^{i+1})}$$

Walshaw [12] proposed $\gamma = \sqrt{7/4}$ for keeping coordinates fixed and reducing spring's natural length $K^i$ in spring-electrical model.

$$K^i = \frac{K^{i+1}}{\gamma}$$

He deduced this equation from a test on a graph with 4 nodes. However, in the proposed method, coordinates could be fixed and determined $K^i$ and the desired length of an edge in force-directed algorithm would be equal to average edge length in the primary layout of $G^i$.

## 2.3  Another Model for Multi-Level Approach

In proposed models for multi-level approach, each large graph $G^{i+1}$ is the product of the coarsening process on previous graph $G^i$ and therefore it is $G^{i+1}$ graph that directly determine the primary layout of $G^i$ in refinement phase. Besides this model, another approach is proposed for creating a sequence of large graphs. In this method which is shown in Fig. 10., each large graph $G^i$ is the product of the coarsening process over primary graph $G = G^0$, the amount of coarsening is controlled by variable $i$ and its growth will lead to a larger graph. For example, coarsening could be performed in a way that each $G^i$ graph has $\frac{1}{2^i}$ nodes with respect to the primary graph. In this approach, coarsening process would be different from introduced method in previous section and should be done in a manner that the size of produced graph can be controlled by variable $i$. The refinement phase has been performed in a different process too. In order to specify the primary layout of $G^i$ while $G^{i+1}$ is at its final layout, at first step the nodes of $G$ are located in the coarsening process based on $G^{i+1}$ layout. Then this layout is used to determine the primary layout of $G^i$. This approach facilitates the implementation of graph partitioning methods for larger graphs which could be done with higher accuracy and based on different parameters. In addition, the coarsening process in each level would be independence from previous graphs and a weak coarsening cannot have a negative impact on all results after itself.

```
pAlgorithm MultilevelLayout_2 (G)
    Initialization;
    Graph G⁰ = G;
    i = 0;
    while (|Vⁱ| ≥ threshold)
        Graph G^{i+1} = CoarsenGraph(G, |V|/2^{i+1});
        i = i + 1;
    RandomLayout (Gⁱ);
    while (i ≥ 0)
        ComputeLayout(Gⁱ);
        if (i ≥ 1)
            InterpolatePositions (G ← Gⁱ)
            InterpolateInitialPositions(G^{i-1} ← G);
        i = i - 1;
    return G⁰;
```

**Fig. 10**. Proposed multi-level algorithm

*2.3.1 Graph Coarsening*

In this new model, the coarsening process is done by means of graph's partitioning algorithms. For producing $G^i$, the primary graph divides $G$ to $N^i$ separate parts (which is approximately equal to $\frac{|V|}{2^i}$) then $G$ is generated by combining the nodes in each partition and eliminating the multiple edges. Specifying the weights for nodes and edges can be done through a technique which is similar to previously discussed methods. For graph partitioning there is a wide range of techniques, in this study spectral clustering [81] is implemented and an algorithm which is based on k-centers technique is presented.

*2.3.2 Refinement Phase*

Extending the final layout of $G^{i+1}$ to the primary layout of $G^i$ is done in two phases. Firstly, the nodes of $G$ are located based on the coarsening algorithm and the $G^{i+1}$ final layout. Then coordinates of $G$ nodes are used for identifying the primary layout of $G^i$. In the proposed model the extending process is done in following steps:

- **Identifying the positions of $G$**: Two variables $(R, O)$ are allocated to every partition of $G$ (resulted from partitioning algorithm for creating $G^{i+1}$). These two variables limit the area that partition's nodes can place themselves in the process of the extension to a circle with center of $O$ and radius of $R$. The value for $O$ is determined based on corresponding node in $G^{i+1}$ and the value of $R$ is adjusted based on lowest distance between this node and the other nodes in $G^{i+1}$, then each node of partition place in a random position in this area. The resulted layout is enhanced by a force-directed algorithm with very few iterations that consider the displacement limitation of different nodes.

- **Identifying the primary layout of $G^i$**: The position of each node in $G^i$ is regulated based on average position of the nodes in the corresponding partition of $G$.

After applying extension process for identifying the primary layout of $G^i$, this layout would be refined with force-directed algorithm.

## 2.4. Fuzzy Multi-Level Model

In this paper a new multilevel approach is introduced which acts based on the fuzzy clustering algorithms.

*2.4.1 Fuzzy Partitioning*

Fuzzy partitioning methods allow nodes to belong to many partitions with different membership degrees simultaneously. In many situations the fuzzy partitioning is more reasonable than the normal partitioning. The nodes, which have similar connections with different partitions, would not be forced to belong to the one of these partitions solely; thus, the allocated membership degrees that are series of numbers between 0 and 1 define their partial membership. The fuzzy partitioning of $G = (V, E)$ to $c$ parts with partition matrix $U = [\mu_{ik}]_{c \times N}$ is shown. The $i_{th}$ row in this matrix shows the membership values for the nodes in the $i_{th}$ partition. The conditions for fuzzy partitioning are demonstrated through following sets of conditions:

$$\mu_{ik} \in [0,1], \quad 1 \leq i \leq c, \quad 1 \leq k \leq N,$$

$$\sum_{i=1}^{c} \mu_{ik} = 1, \quad 1 \leq k \leq N,$$

$$0 < \sum_{k=1}^{N} \mu_{ik} < N, \quad 1 \leq i \leq c.$$

The second equation implies that the sum of the items in each column would be equal to 1 and it equivalent to this fact that total memberships of each node in different partitions would be equal to 1, limiting $\mu_{ik}$ values to 0 and 1 would lead to a normal partitioning.

*2.4.2 Fuzzy Multi-Level Model*

In the fuzzy multi-level model, the process of generating large graph $G^{i+1}$ from small graph $G^i$ divides into two steps. First step is fuzzy partitioning algorithm which creates the partition matrix $U = [\mu_{ik}]_{c \times N_i}$ from graph $G^i$ with $c \approx \frac{N_i}{2}$. Then $G^{i+1}$ with $c$ nodes is created and its edges and weights of its nodes and edges are determined based on $U$ matrix. Refinement phase is also done in two steps. First, the position of each node in $G^i$ is calculated based on joined clusters and position of corresponding nodes in $G^{i+1}$, then this position is enhanced by defining a movement area for the node which is cannot pass over and applying a force-directed algorithm with very few iterations. The refinement phase for $G^i$ layout is executed like previous methods with force-directed algorithm.

- Creating Partition Matrix

One of the most important fuzzy clustering techniques is c-means fuzzy clustering technique [1004]. Using this method for partitioning requires the presentation of the nodes in a vector with real values. This vector should be calculated in a way that the Euclidean distance of vectors represent the distance of their corresponding nodes in graph. In other words, lower distance between vectors is equal to more connectivity and higher distances show weaker connections. Calculation of such vector in $R^2$ space can be represent a solution for graph drawing problem. We have two proposes for creating such vector, the first purpose is the calculation of distance-vector between nodes which is impractical for large graphs because of their great dimensions (which is equal to the number of nodes in graph). Second purpose is a vector based on spectral clustering algorithm which is applied for utilization of simple c-means algorithm.

In order to improve the efficiency of the fuzzy multi-level model, a quick method is required to estimate the partition matrix of the graph. To do so, a method is implemented which is similar to technique which was introduced in [32]. In this method, partition matrix is determined in two steps. In the first step, sub-sets of the nodes are selected as the center of the targeted clusters. Then several rules would be defined that help for the identification of the relationships between external node and central nodes and their coefficients (membership degree). In order to select the center of targeted clusters two approaches are used. The first one is similar to the technique which was mentioned in [32] and the other one is a simplified version of k-centers algorithm which has following steps:

- Firstly, one node is randomly selected, then a group of targeted nodes are equalized with it.

- Secondly, in each phase a node which is at maximum distance from targeted nodes is selected and added to the list of target nodes until reaching to a certain number of nodes.

The central nodes of each cluster only belong to the same cluster. For the estimation of the membership degree in rest of the nodes a technique like [32] is implemented.
- Producing the coarse graph based on the Partition Matrix

Process of creating the large graph $G^{i+1}$ consists of following steps:
- The number of nodes in $G^{i+1}$ would be equal to the number of partitions in $G$ (the number of columns in $U$ matrix)

- The weight of each node in $G^{i+1}$ should reflect the sum of nodes weight in corresponding partition. This weight would be equal to the sum of each node's weight multiplied in its membership degree.

$$\boldsymbol{\omega}_i^{i+1} = \sum_{k=1}^{N} \boldsymbol{\omega}_k^i \cdot \boldsymbol{U}_{ik}$$

- The weight of an edge $(p, q)$ in $G^{i+1}$ should represent the weights of all connected edges between corresponding partitions $(p, q)$ in $G^i$.

$$\omega_{pq}^{i+1} = \sum_{k,l \in V(G^i)} U_{pk} \cdot \omega_{kl}^i \cdot U_{ql}$$

- The proportion of the number of edges to nodes in the resulted large graph, often is significantly greater than primary graph. This might generate some problems in the optimization process. In [32] some methods had been proposed to deal with this issue that had been used too.

- Refinement Phase

Each node of $G^i$ belongs to a group of partitions with different membership degrees. Each of these partitions are related to a corresponding node in $G^{i+1}$. Based on this fact, the primary position of the nodes $G^i$ and $X^i$ are calculated through following equation with respect to the location of $G^{i+1}$ and $X^{i+1}$:
$$X^i = U^T \times X^{i+1}$$

In this equation $U^T$ is the transposed matrix of $U$, after the estimation, by using a method which was described in previous section, the position of $G^i$ nodes had been enhanced locally through force-directed algorithm.

## 3. Conclusion

The current research investigated general graphs and used the combination of force-directed algorithms and multi-level algorithms. Force-directed approach with respect to three independent partial problems is investigated as follows:

1. Defining force rules which measure aesthetic metrics

2. Calculation of forces for acquiring the negative gradient of an implicit energy function

3. Applying numerical optimization for finding local minimum of this function

The main achievement of this research with respect to force-directed approach is describing it for general graph drawing modeling as a numerical optimization problem; consequently, it can use rich knowledge which is proposed by numerical optimization as an established system.

Furthermore, multi-level approach is a useful intuitive tool which is applied to overcome the local minimums in standard force-directed algorithms. Two new models are introduced in field of multi-level approaches beside current models. The first model uses from the graph partitioning methods for determination of larger graphs during drawing process and the second model is fuzzy multi-level approach which is based on fuzzy partitioning algorithms. The first method applies partitioning techniques in a different manner for selection of larger graphs and make it possible to have a higher degree of control over mid-level large graphs. The other model is fuzzy multi-level model which acts based on fuzzy clustering algorithms and it seems like it is possible to achieve better results through its development and enhancement. The observations show the relative superiority of this model with respect to other multi-level approaches and in the viewpoint, this is the direct result of this fact that large graphs which are created through fuzzy partitioning usually have a more efficient representation of high-level graph structures in comparison with other coarsening methods. It should be mentioned that the main focus of this research is the quality of drawing results. Also, it should be noted that we have designed a program for the simulation of described algorithms. Therefore, the results of applying algorithms on some of selected graphs are shown in appendix A.

**Appendix A**

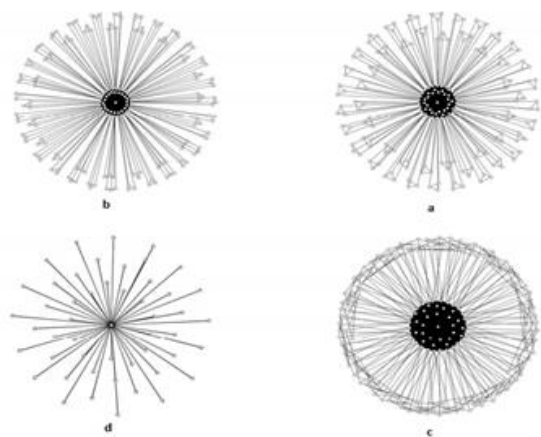

**Fig. A.1.** Results of drawing for a: head to head attraction and repulsion forces, b: binary stress function, c: spring force, d: Linlog function

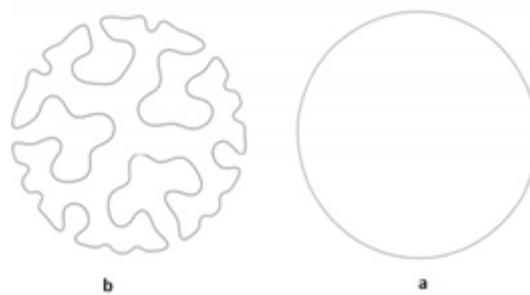

**Fig. A.2.** Results for a: stress function, b: binary stress function

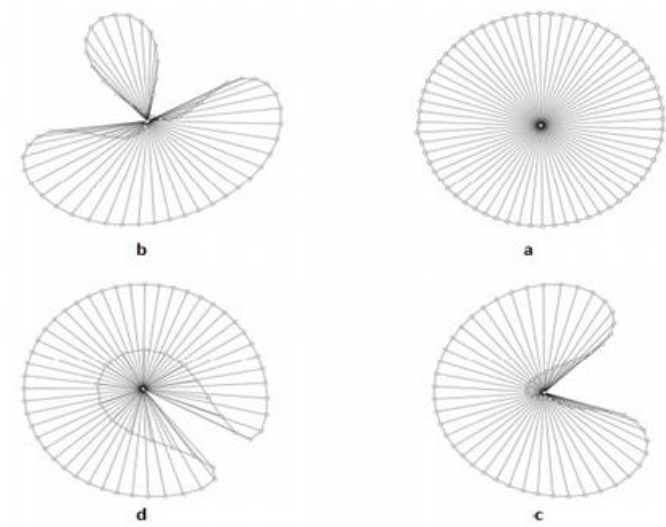

**Fig. A.3.** Drawing results for a: primary graph, b: Linlog function, c: Linlog function, d: binary stress function

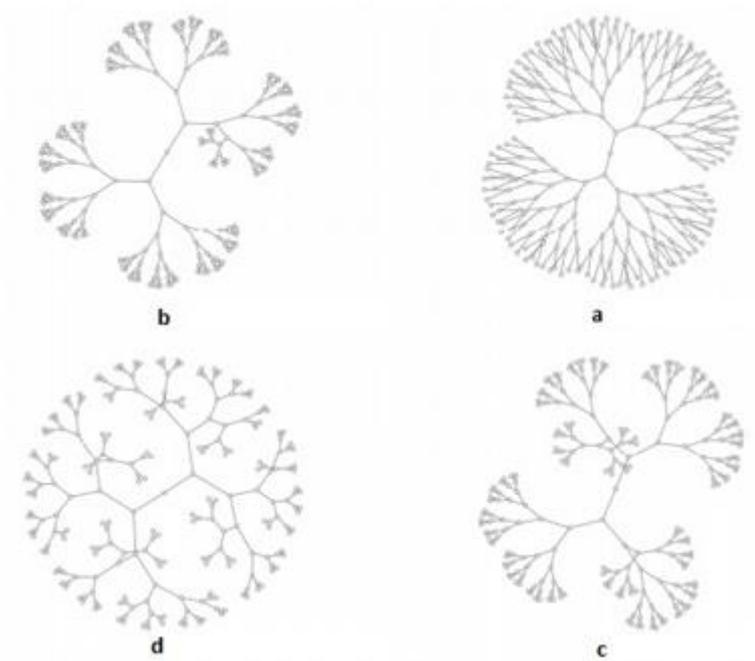

**Fig. A.4.** Drawing results for a: stress function, b: head to head attraction and repulsion, c: head to head attraction and repulsion between nodes + repulsion force between edge and node, d: binary stress function

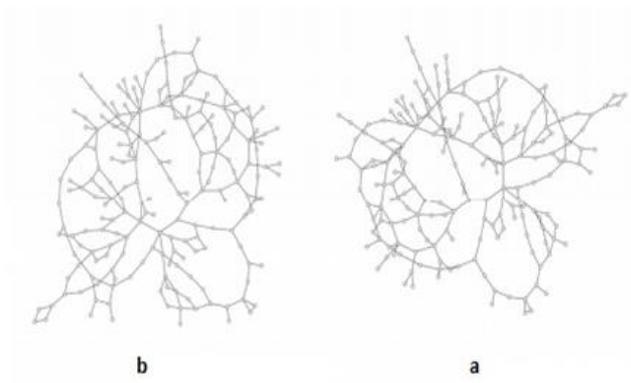

**Fig. A.5.** Drawing result for a: attraction and repulsion force between nodes + stress function, b: head to head attraction and repulsion + stress function + repulsion between node and edge

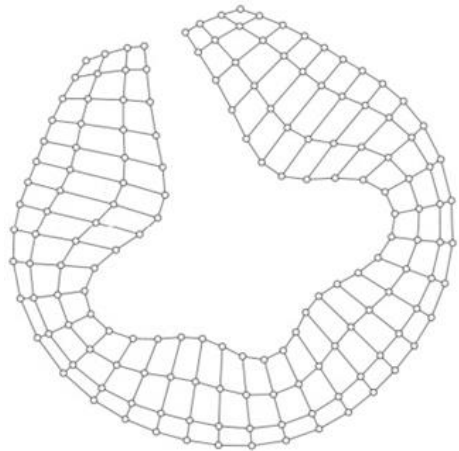

**Fig. A.6.** Drawing result for binary stress function

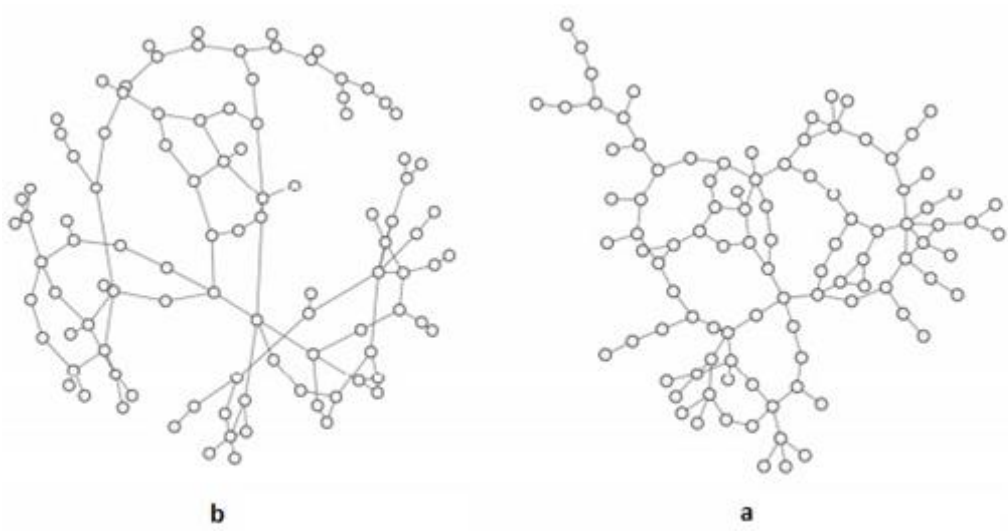

**Fig. A.7.** Drawing results for a: stress function, b: binary stress function

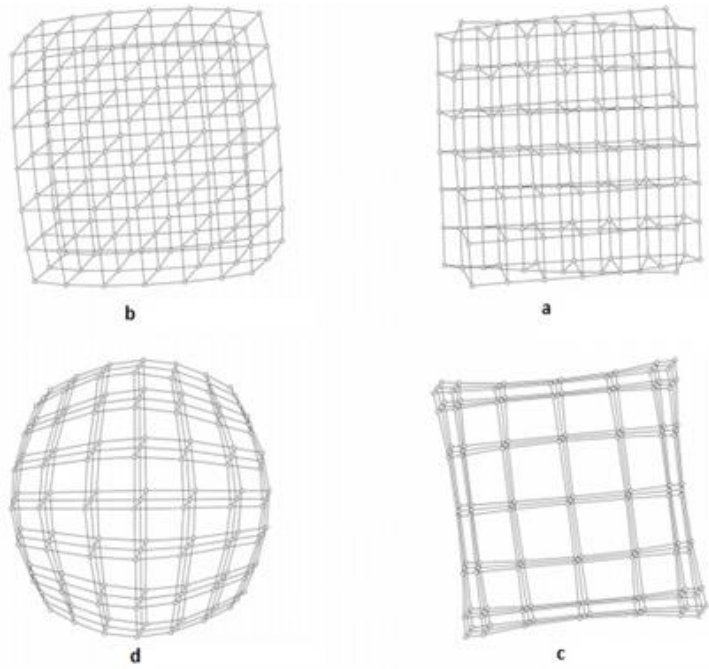

**Fig. A.8**. Drawing results for **a.** stress function, **b.** head to head attraction and repulsion, **c**. Linlog function, **d.** binary stress function

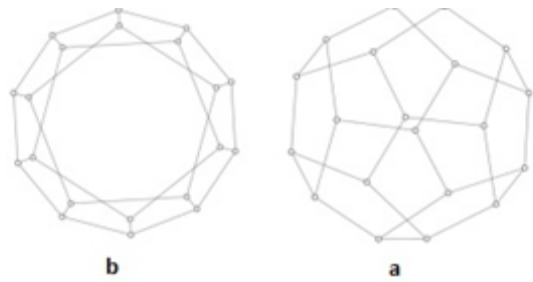

**Fig. A.9.** Drawing results for graph with 12 components **a.** stress function, **b.** Linlog function

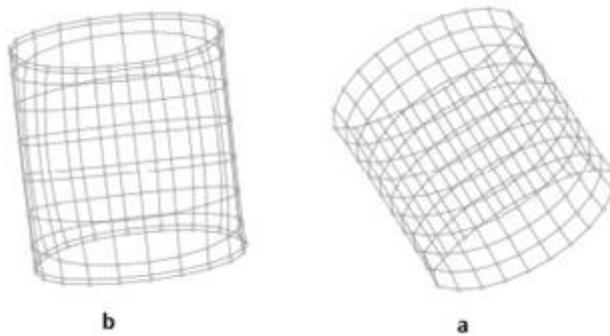

**Fig. A.10.** Drawing results for a**.** stress function, **b.** Linlog + binary stress function

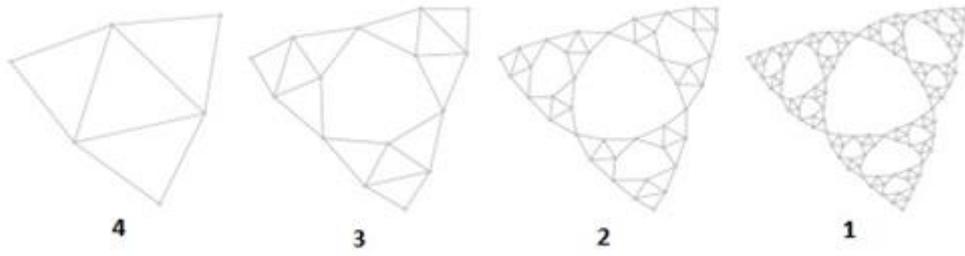

**Fig. A.11.** Large graphs produced by fuzzy enlargement algorithm

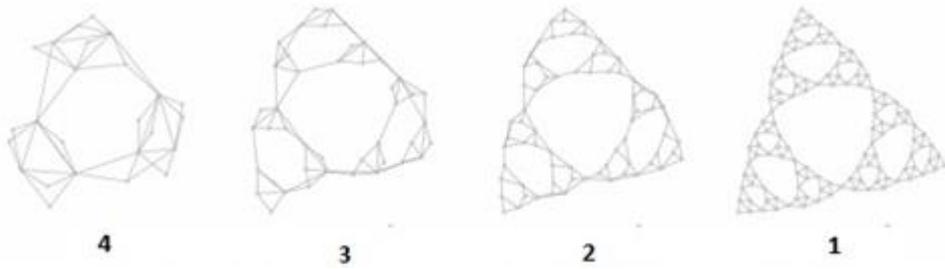

**Fig. A.12.** Large graphs produced by EC enlargement algorithm

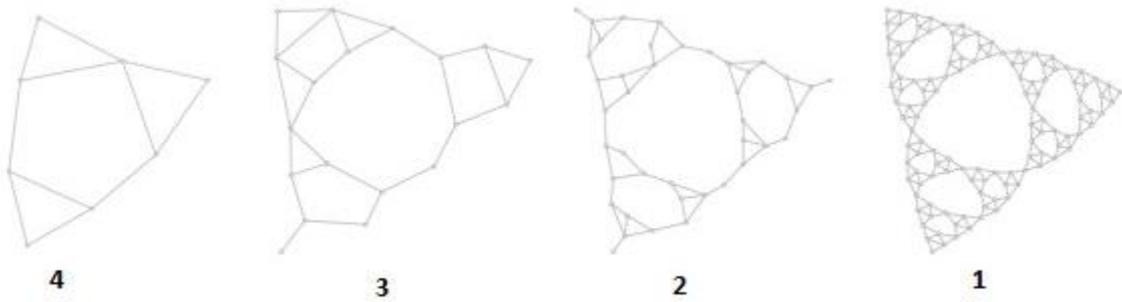

**Fig. A.13.** Large graphs produced by MIVS enlargement algorithm

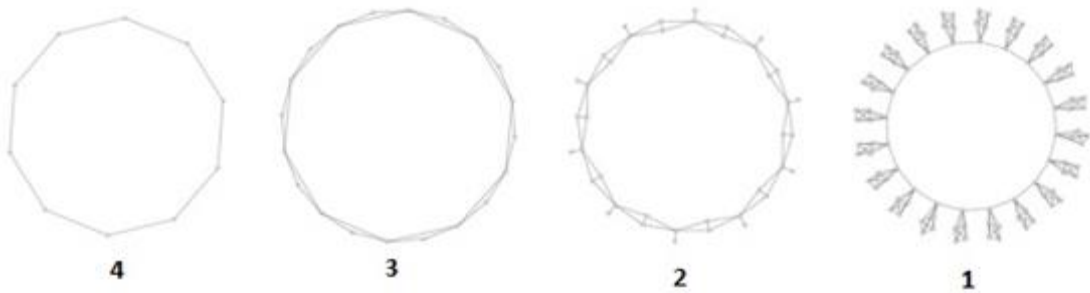

**Fig. A.14.** Large graphs produced by fuzzy enlargement algorithm

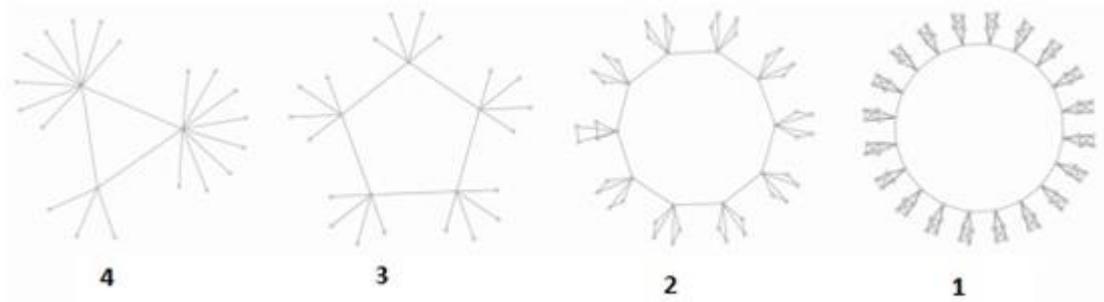

**Fig. A.15.** Large graphs produced by EC enlargement algorithm

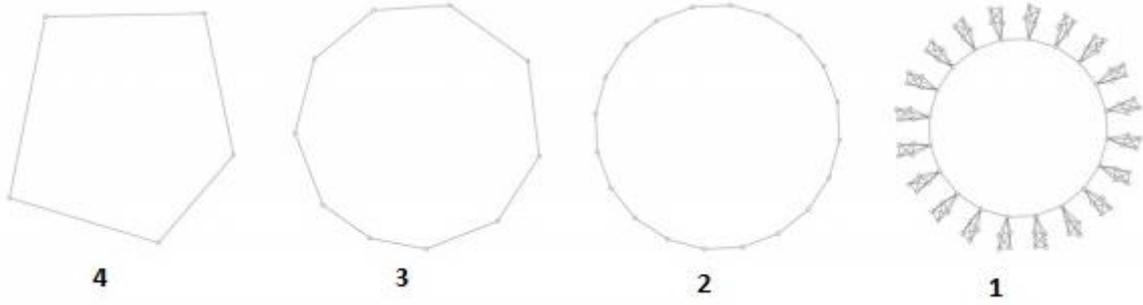

**Fig. A.16.** Large graphs produced by MIVS enlargement algorithm

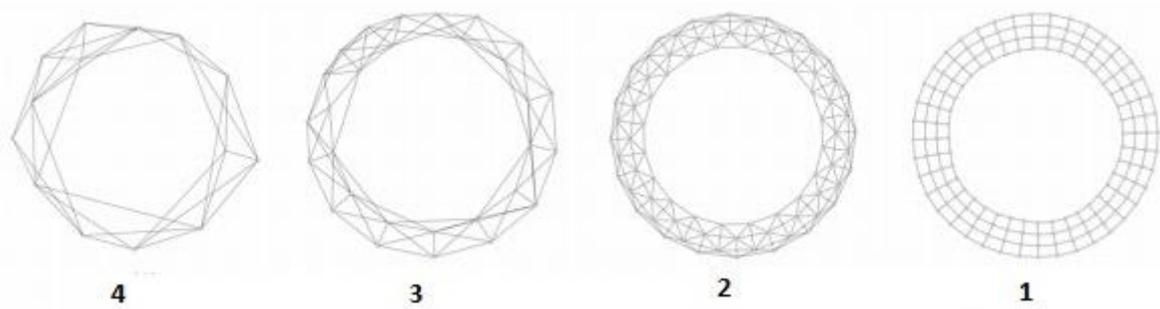

**Fig. A.17.** Large graphs produced by fuzzy enlargement algorithm

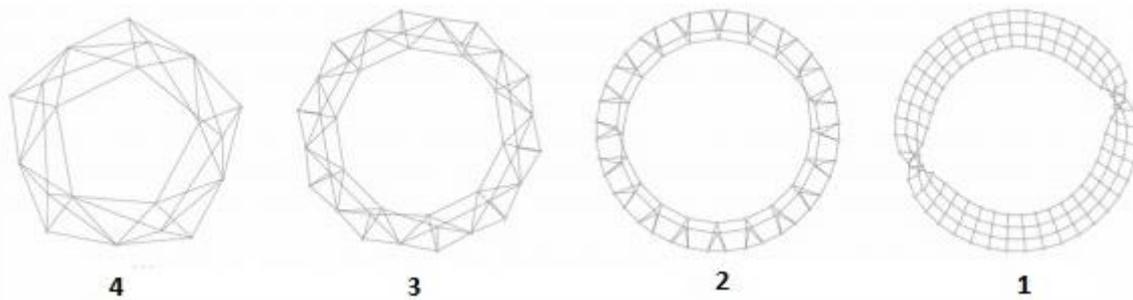

**Fig. A.18.** Large graphs produced by EC enlargement algorithm

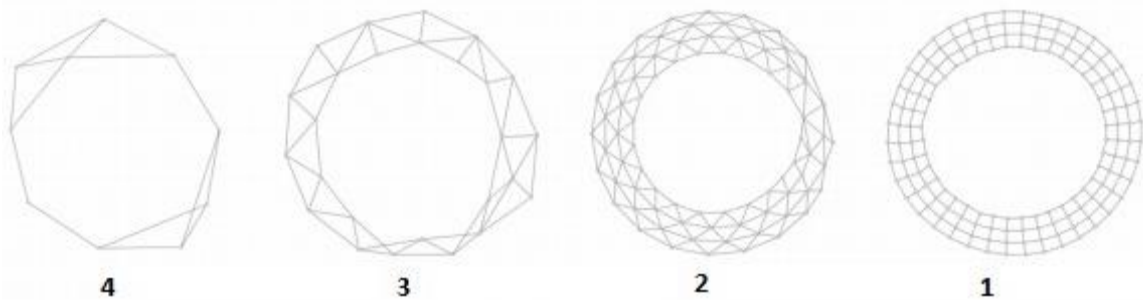

**Fig. A.19.** Large graphs produced by MIVS enlargement algorithm